\documentclass[onecolumn, doublespace, 12pt fonts]{aastex6}

\shorttitle{Late-Time X-ray Flares}
\shortauthors{Mu et al.}

\begin{document}

\title{Central Engine of Late-Time X-ray Flares with Internal Origin}

\author{Hui-Jun Mu\altaffilmark{1}, Wei-Min Gu\altaffilmark{1},
Shu-Jin Hou\altaffilmark{2}, Tong Liu\altaffilmark{1,4},
Da-Bin Lin\altaffilmark{3}, Tuan Yi\altaffilmark{1},
En-Wei Liang\altaffilmark{3}, and Ju-Fu Lu\altaffilmark{1}}

\altaffiltext{1}{Department of Astronomy and Institute of Theoretical Physics
and Astrophysics, Xiamen University, Xiamen, Fujian 361005, China;
guwm@xmu.edu.cn}
\altaffiltext{2}{College of Physics and Electronic Engineering,
Nanyang Normal University, Nanyang, Henan 473061, China}
\altaffiltext{3}{GXU-NAOC Center for Astrophysics and Space Sciences,
Department of Physics, Guangxi University, Nanning 530004, China}
\altaffiltext{4}{Department of Physics and Astronomy, University of Nevada,
Las Vegas, NV 89154, USA}

\begin{abstract}
This work focuses on a sample of seven extremely late-time X-ray flares
with peak time $t_{\rm p} > 10^4 {\rm s}$, among which two flares
can be confirmed as the late-time activity of central engine.
The main purpose is to investigate the mechanism of
such late-time flares based on the internal origin assumption.
In the hyper-accreting black hole (BH) scenario, we study the possibility
of two well-known mechanisms as the central engine to power
such X-ray flares, i.e., the neutrino-antineutrino annihilation and
the Blandford-Znajek (BZ) process.
Our results show that the annihilation luminosity is far below
the observational data. Thus, the annihilation mechanism cannot
account for such late-time flares.
For the BZ process, if the role of outflows is taken
into consideration, the inflow mass rate near the horizon
will be quite low such that the magnetic field will probably be too weak
to power the observed X-ray flares.
We therefore argue that, for the late-time flares with internal origin,
the central engine is unlikely to be associated with BHs.
On the contrary, a fast rotating neutron star with strong bipolar
magnetic fields may be responsible for such flares.
\end{abstract}


\section{Introduction}\label{sec:Introduction}

In the past twenty years, great progress has been achieved on the
understandings of gamma-ray bursts (GRBs).
In particular, Swift has opened a new window to understand the nature
of GRB phenomenon
(e.g., \citet{Meszaros2006,Zhang2007,Liang2010}).
The onboard X-Ray Telescope (XRT; \citealp{Burrows2005a}) opened an
exciting era for GRB researches.
It has established a large sample of X-ray light curves from tens
of seconds to days, sometimes even months (e.g., GRB 060729;
\citealp{Grupe2006}).
It is interesting to find that X-ray flares are common in GRBs,
which occur well after
the initial prompt emission \citep{Romano2006,Falcone2007,
Chincarini2007,Margutti2010,Margutti2011,Bernardini2011}.
X-ray flares have been observed both in long and short GRBs
\citep{Romano2006,Falcone2006,Campana2006,Margutti2011}.
Based on the observations from Swift/XRT,
four power-law light-curve segments together with a flaring component
are identified in the X-ray afterglow phase
\citep{Zhang2006,Nousek2006,OBrien2006}.
The temporal analysis and spectral property
suggest that the X-ray flare is from a distinct emission mechanism,
since the temporal behavior of flares is quite similar to
the prompt emission pulses, whereas different from the other four components
in the canonical light-curves.
Thus, X-ray flares may have a common physical origin
as the prompt pulses \citep{Burrows2005b,Falcone2006,Falcone2007,
Liang2006,Nousek2006,Zhang2006,Chincarini2007,Chincarini2010,
Wu2013,Hou2014a,Yi2015},
and are probably related to the late time activity of the central engine
\citep{Romano2006,Bernardini2011}.

The central engine of GRBs remains an open question \citep{Zhang2011}.
A hyper-accreting stellar-mass black hole (BH)
or a millisecond magnetar \citep{Usov1992,Dai1998,Zhang2001,Metzger2011}
is usually invoked as possible GRB central engine.
In the BH hyper-accretion scenario, the photons generated in the accretion
flow can hardly escape due to the extremely large optical depth. On the
contrary, a large amount of neutrinos can escape from the flow and therefore
the neutrino cooling may be the most important mechanism to balance the
viscous heating. Such a flow is named as the
neutrino-dominated accretion flow (NDAF).
The structure and radiation of an NDAF has been
extensively studied (e.g., \citet{Popham1999,Di2002,Gu2006,Kawanaka2007,
Liu2007,Liu2013,Liu2014,Lei2009,Zalamea2011,Pan2012,Janiuk2013,Xue2013,Cao2014}.
In the hyper-accretion scenario, the relativistic jet may be powered
by the following two mechanisms.
The first one is related to the
annihilation of neutrino-antineutrino pairs.
Such an annihilation process was previously investigated by several works
(e.g., \citet{Popham1999,Di2002,Gu2006,Birkl2007,Liu2007,Xue2013,Liu2016}.
The second one is related to the Blandford-Znajek (BZ) process
\citep{Blandford1977}, which can effectively extract the rotational energy of
the central BH through large-scale magnetic fields.

The physical origin of X-ray flares remains mysterious,
including internal dissipation and external shock mechanisms.
\citet{Falcone2007} showed that many X-ray flares are from late-time
activity of the internal engine that spawned the initial GRB, not from
an afterglow-related effect.
Moreover, \citet{Chincarini2010} and \citet{Margutti2010}
made analyses of the flare temporal and spectral properties of
a large sample of early-time flares and of a subsample of bright flares,
which revealed close similarities between them and the prompt emission pulses,
and therefore pointing to an internal origin.
\citet{Margutti2011} investigated the relation between flares and
continuum emission, and suggested the variability to be established
as a consequence of different kinds of instabilities.
On the other hand, from the theoretical view, \citet{Ioka2005} proposed
a criterion to separate the internal and external origin of flares.
\citet{Curran2008} concluded that the late-time flares
($t_{\rm p}\ga 10^{4}{\rm s}$) are not different from the early-time ones,
where the majority of the flares can be explained by either internal
or external shock.
However, due to the small number of flares (a sample of 7 GRBs),
the conclusion may require further investigation.
Moreover, \citet{Bernardini2011} focused on the late-time flares
($t_{\rm p} \ga 10^{3}$~{\rm s}) of a larger sample than \citet{Curran2008}
and found that a large fraction of late-time flares are also
compatible with afterglow variability.
In addition, \citet{Lazzati2007} showed internal dissipation
and external shock mechanisms of X-ray flares,
and concluded that at least a sizable fraction of the flares
cannot be related to the external shock mechanism, since
external shock flares evolve on much longer timescales than observed.
Then, some late-time flares in our sample may be related to late-time
central engine activity rather than a slower outflow produced
simultaneously with the prompt emission.
Moreover, the steep decay of X-ray flares is more likely to
originate from the internal dissipation (e.g., \citep{Kumar2000}).
In the present work, we will adopt two criteria to examine the internal or
external origin of the flares in our sample, and then study the possible
mechanisms for those flares probably related to internal origin.

Several mechanisms and models were proposed to explain the episodic
phenomenon of X-ray flares
\citealp{King2005,Dai2006,Meszaros2006,Perna2006,Lazzati2008,Lee2009,
Lazzati2011,Yuan2012,Luo2013,Hou2014b}.
According to the internal origin of X-ray flares, the central engine
that powers the prompt gamma-ray emission also powers the X-ray flares.
Thus, the long-duration flares require the long-lasting activity
of the central engine.
For the neutrino annihilation mechanism, as pointed out by \citet{Luo2013},
although such a mechanism may work well for the central engine of the
gamma-ray emission, it may encounter difficulty in interpreting the
X-ray flares. By considering a possible magnetic coupling between
the inner disk and the central BH, \citet{Luo2013} showed that
the annihilation mechanism can also work for the X-ray flares with duration
$\tau \la 100{\rm s}$.
However, the annihilation mechanism is unlikely to be
responsible for those long flares with duration $\ga 1000{\rm s}$,
even the role of magnetic coupling is included.
On the other hand, according to the analyses of \citet{Luo2013},
the BZ mechanism may work well even for the long duration flares.
However, outflows were not taken into consideration
in \citet{Luo2013}, which can be of importance particularly for
relatively low accretion rates where the neutrino cooling is inefficient.

There is a positive correlation for the X-ray flares between the duration
$\Delta t$ and the peak time $t_{\rm p}$ (e.g., \citealp{Margutti2010,Yi2016} ).
In the present work, we will focus on
the extremely late-time X-ray flares with $t_{\rm p} > 10^4$~s
and study the corresponding central engines.
The remainder of this paper is organized as follows.
Our sample and data analyses are presented in \S~2.
The different mechanisms for the central engine
are investigated in \S~3.
Conclusions and discussion are made in \S~4.

\section{Sample and data analysis}

We present an extensive temporal analysis for the X-ray
afterglow observed by Swift/XRT,
and consider all X-ray afterglow light curves of GRBs detected by
Swift/XRT during 11 observation years
(from 2005 to 2015) in this repository.
The XRT flux lightcurve data and redshift were downloaded from
the website \url{http://www.swift.ac.uk/xrt curves/}
\citep{Evans2007,Evans2009}.
We examined visually all the light curves and searched for bright
X-ray flares with extremely late-time, i.e. $F_{\rm p}> 3F$ and
 $t_{\rm p} > 10^4 {\rm s}$,
where $F_{\rm p}$ and $F$ are the peak flux of the flare and
the flux of the underlying continuum at $t_{\rm p}$, respectively.
We obtain a sample of seven flares from seven GRBs, among which
three GRBs have redshift measurements.
The spectral analyses for the steep decay segments are
based on \url{http://www.swift.ac.uk/xrt spectra/addspec.php/},
which is performed by a power-law spectral model.
The spectral analyses results, i.e., the values of the spectral index
in the decay phase $\beta$, are shown in Table~1.

\subsection{Light-curve fitting}

In order to estimate the duration and luminosity of flares,
a smooth broken power-law function \citep{Liang2007,Li2012,Yi2016}:
\begin{equation}\label{BPL}
F_{\rm{t,f}}=F_{0}\left[(\frac{t}{t_{\rm{b}}})^{\alpha_{1}\omega}+(\frac{t}
{t_{\rm{b}}})^{\alpha_{2}\omega}\right]^{-\frac{1}{\omega}},
\end{equation}
and a power-law function:
\begin{equation}\label{PL}
F_{\rm{t,a}}=F_{\rm{0,a}}t^{-\alpha}\ ,
\end{equation}
are used to fit the light curves of flares and the underlying
continuum, respectively.
Here, $\alpha_{1}$ ($\alpha_{2}$) is the rise (decay) index of X-ray flare,
$t_{\rm{b}}$ is the break time,
$\alpha$ is the decay index of the underlying afterglow component,
and $\omega=3$ is used to depict the sharpness around peak flux
in the flare light-curve.
We would point out that, for the late-time X-ray flares in our sample,
it is possible that some of them are a superposition of many shorter flares.
In such case, the duration of X-ray flares may be overestimated by
a factor of a few. Our main concern in this work is the duration of
the flare emission episode. Thus, whether a long-duration flare or many
shorter flares may not have essential influence on our analysis.

\begin{figure}
\figurenum{1}
\plotone{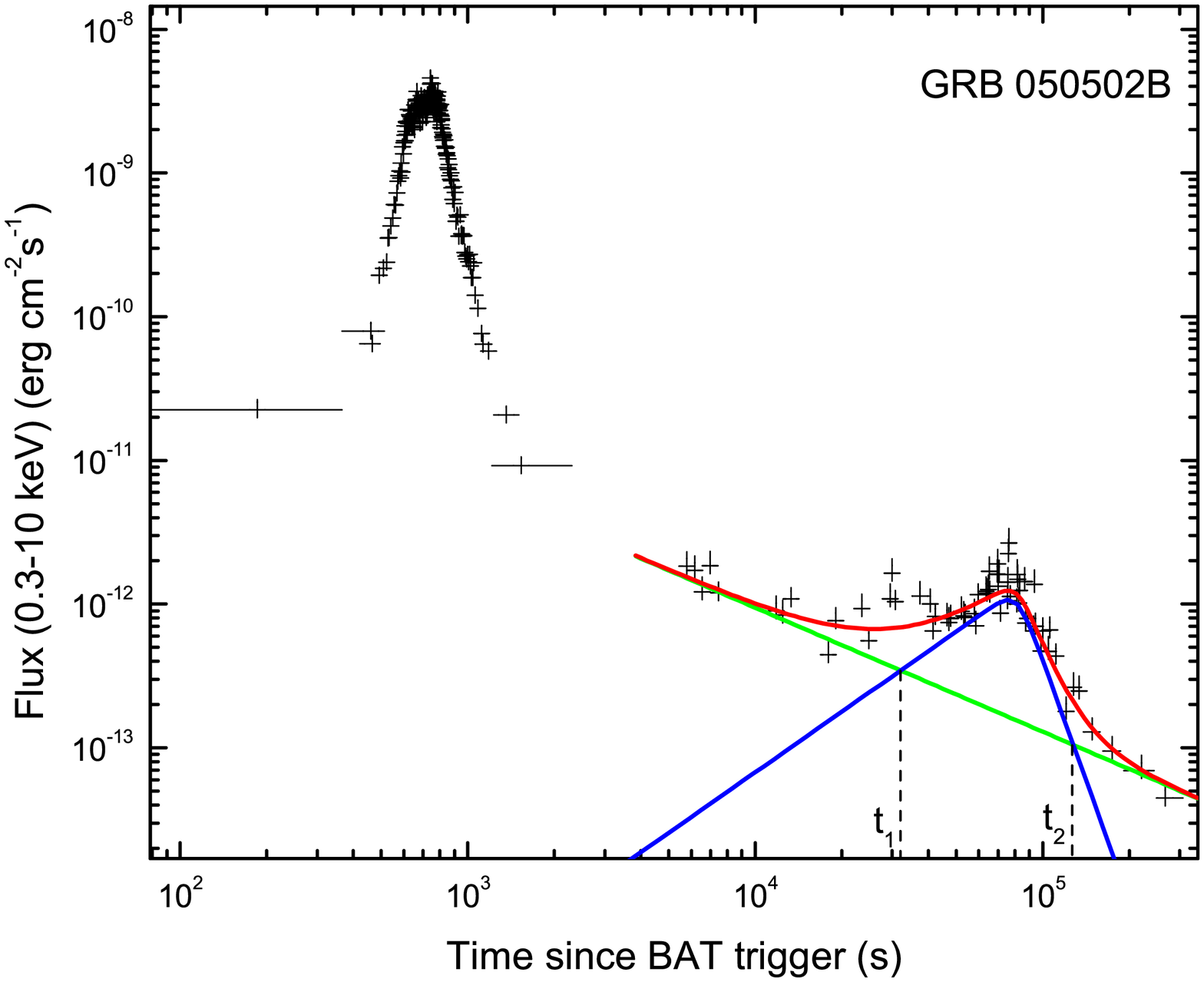}
\caption{
The best fitting for the late-time
X-ray afterglow light curve of GRB 050502B
(red curve). The blue curve and the green line show the best fitting of
the late-time flare and the underlying continuum, respectively.
The first flare is not considered in this fitting.
}
\label{F:050502B}
\end{figure}

\begin{figure}
\figurenum{2}
\centering
\plotone{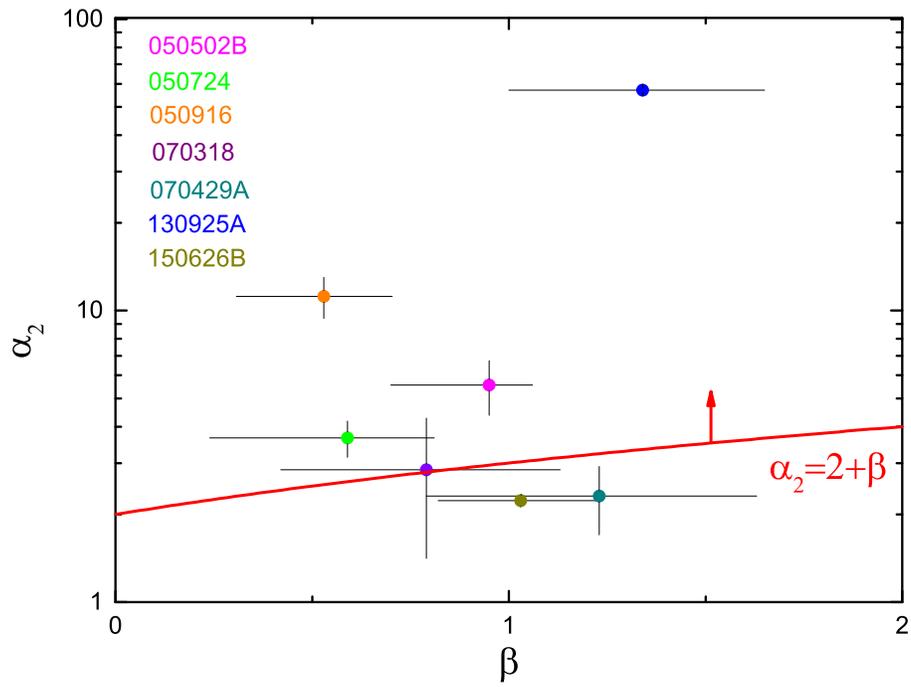}
\caption{
A comparison of the seven late-time flares in our sample with
the criterion of internal origin ``$\alpha_2 > 2+\beta$".
}
\label{F:alpha and beta}
\end{figure}

Each X-ray flare from our sample is fitted by a smooth broken power-law
function as shown by Equation~(\ref{BPL}).
The peak time $t_{\rm{p}}$ can be calculated as
\begin{equation}\label{tp}
t_{\rm p} = t_{\rm b}(-\frac{\alpha_{1}}{\alpha_{2}})
^{\frac{1}{(\alpha_{2}-\alpha_{1})\omega}}\ .
\end{equation}
The main fitting results are listed in Table~1.
As an example, Figure~\ref{F:050502B} illustrates the fitting procedure
of GRB 050502B.
Here, we define the duration $\Delta t$ as the full width at half maximum
(FWHM) of the X-ray flares, and
\begin{equation}
\Delta t_{\rm res}=\Delta t/(1+z)
\end{equation}
is the duration in the rest frame.
By setting the zero time $T_0$ at the GRB trigger time,
the flares formed in the external shock process has a maximum
decay slope $\alpha_2 = 2+\beta$,
where $\alpha_2$ and $\beta$ are the
temporal decay index and spectral index in the decay phase, respectively.
Any decay with a slope steeper than $2+\beta$, i.e. $\alpha_2 > 2+\beta$, may
indicate the internal origin of flares \citep{Kumar2000,Liang2006}.
Then, we compare the values of $\alpha_2$ and $2+\beta$ for our seven
late-time X-ray flares in Figure~\ref{F:alpha and beta}.
It is seen that four flares in our sample locate well above the red solid
line, which means that the four flares satisfy the criterion
``$\alpha_2 > 2+\beta$", and therefore are likely to be internal origin.
In the remainder, we will focus on these four flares.

\subsection{Isotropic luminosity and energy}

The isotropic energy $E_{\rm{X,iso}}$ of a single X-ray flare in XRT
energy range is calculated by
\begin{equation}\label{Ex,iso}
E_{\rm{X,iso}}=\frac{4\pi D_{\rm{L}}^{2}S_{\rm{F}}}{1+z} \ ,
\end{equation}
where $D_{\rm{L}}$ is the distance of GRB with respect to the observer, and
\begin{equation}\label{Fluence}
S_{\rm{F}}=\int^{10t_{2}}_{0.1t_{1}}F_{0}\left[(\frac{t}{t_{\rm{b}}})
^{\alpha_{1}\omega}+(\frac{t}{t_{\rm{b}}})^{\alpha_{2}\omega}\right]
^{-\frac{1}{\omega}} \ dt \ ,
\end{equation}
is the energy fluence in the energy range of Swift/XRT (i.e., 0.3-10~\rm{keV}).
Here, $t_{1}$ and $t_{2}$ ($t_1 < t_2$)
can be derived by the two cross points of the
two curves corresponding to Equations~(\ref{BPL}) and (\ref{PL})
(see \citet{Falcone2007} and \citet{Yi2016}).
We choose a sufficiently large time interval, i.e., from 0.1$t_1$ to 10$t_2$,
for the integration of Equation~(\ref{BPL}).
The isotropic luminosity $L_{\rm{X,iso}}$ of a single X-ray flare
is expressed as
\begin{equation}\label{Lx,iso}
L_{\rm{X,iso}}=\frac{(1+z)E_{\rm{X,iso}}}{\Delta t}.
\end{equation}
In addition, the anisotropic effects of the jet radiation should be taken
into account. We adopt $1-\cos\theta_{\rm jet} \approx 0.1$, i.e.,
$\theta_{\rm jet}\sim 0.45$~radian, where $\theta_{\rm jet}$ is the
half-opening angle of the jet related to the flares.
This value is around one order of magnitude
larger than the prompt gamma-ray emission ($\sim 0.05$~radian)
\citep{Zeh2006}.
The reason of adopting such a value is that there may exist a relation
$\gamma \theta_{\rm jet}\sim 20$ \citep{Tchekh2010b}, and
the Lorentz factor $\gamma$ of the jet producing the X-ray flares
may be significantly smaller than that of the gamma-ray emission.
Thus, the collimated-corrected energy
$E_{\rm{X}} = E_{\rm{X,iso}} (1-\cos\theta_{\rm jet}) = 0.1 E_{\rm{X,iso}}$
and luminosity $L_{\rm{X}} = 0.1 L_{\rm{X,iso}}$.
The main results are shown in Table~2,
where the uncertainties of luminosity and energy are given at 1~$\sigma$.
More details about the uncertainties may refer to \citet{Yi2016}.
We would point out that the k-correction is not considered in the
present work.
The luminosity and energy will be higher if the k-correction is taken into
account.
Here, we adopt $z=2$ for those flares without redshift measurements
\citep{Salvaterra2012}.
In order to explore the range of variability of the luminosity and energy
for all possible values of the unknown redshift,
we use $z=0.1$ and 10 as the lower and upper limits, respectively.

\begin{figure}
\figurenum{3}
\centering
\plotone{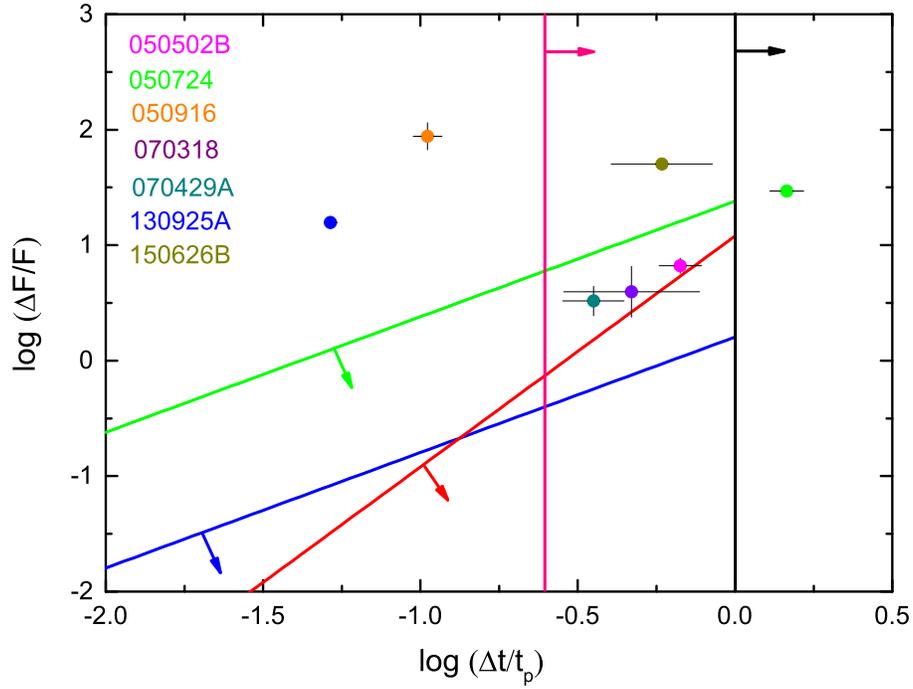}
\caption{
The relationship between the relative variability flux $\Delta F/F$
and the relative variability timescale $\Delta t /t_{\rm p}$ for
the seven late-time X-ray flares in our sample.
The five theoretical solid lines are identical with those in
Figure~6 of \citet{Bernardini2011}, i.e., density fluctuations
on axis (blue line) and off-axis (red line), off-axis multiple regions
density fluctuations (green line), patchy shell model (black line),
and refreshed shocks (pink line).
}
\label{F:Flux and width}
\end{figure}

With the above results, we can examine the relationship between the relative
variability flux $\Delta F/F$ and the relative variability timescale
$\Delta t/{t_{\rm p}}$, where $\Delta F$ is the increase of flux at the
peak time $t_{\rm p}$, and $F$ is the flux of the underlying continuum
at $t_{\rm p}$
(see \citet{Ioka2005} and \citet{Bernardini2011} for details).
In Figure~\ref{F:Flux and width}, we plot exactly the same five theoretical
solid lines as those in Figure~6 of \citet{Bernardini2011}.
The data for our seven late-time X-ray flares are also plotted by
different colors in this figure.
Another criterion to judge the internal origin or not
is to examine the position of a flare in such a figure.
A flare locating in the upper left region (above the green line
and left to the vertical pink line) can be regarded as the internal origin.
On the contrary, the flare locating in other regions may be related to
the external origin.
It is seen that two flares (050916 and 130925A) well locate in the
upper left region, which corresponds to the internal origin.
On the other hand, the other five flares do not locate in this region,
which indicates that the five flares may be related to the external origin.
By combining the results of Figures 2 and 3, we can draw the conclusion
that, the late-time flares of 050916 and 130925A are probably related to
the late-time activity of central engine since both of the two
criteria are satisfied, whereas the flares
of 070318, 070429A, and 150626B are more likely to be the external
origin since neither of the two criteria is matched.
For the rest two flares 050502B and 050724,
however, the physical origin may remain controversial
since the criterion ``$\alpha_2 > 2+\beta$" is matched but the other one
is not.
In particular, for the only short burst 050724,
we noticed that some previous works suggested late-time
activity of central engine \citep{Fan2005,Dai2006}, while some
other works such as \citet{Bernardini2011} argued against internal origin.

In the rest part of this paper, we will focus on the four X-ray
flares which can satisfy at least one criterion,
i.e., 050502B, 050724, 050916, and 130925A. The main purpose is to
investigate the possible central engine, where the internal origin
is our basic assumption for such flares.

\section{Mechanisms of the central engine}

In this section, we focus on the central engine of the sample of
the four late-time X-ray flares based on the energy argument.
As mentioned in Section~1,
there are two well-known mechanisms related to accreting BHs, i.e.,
the neutrino-antineutrino annihilation and the BZ process.
We study these two mechanisms in the first and second subsections,
respectively. In addition, we discuss the possibility of
an NS system as the engine in the third subsection.

\subsection{Neutrino-antineutrino annihilation}

In this subsection, we calculate the neutrino annihilation luminosity based
on previous theoretical formulae.
We assume a typical BH mass $M_{\rm BH}=3M_{\sun}$ and
a spin parameter $a_{*}=0.95$.
We adopt the analytic result, Equation~(22) of \citet{Zalamea2011}
to calculate the annihilation luminosity:
\begin{equation}\label{nunu}
L_{\nu \bar\nu} \approx 1.1 \times {10^{52}}{\mkern 1mu} \chi _{\rm{ms}}^{ - 4.8}(\frac{{{M_{\rm{BH}}}}}{{3{M_ \odot }}}){{\mkern 1mu} ^{ - 3/2}} \times \left\{ {\begin{array}{*{20}{c}}
{0\;\;\;\;\;\;\;\;\;\;\;\;\;\;\;\;\;\;\;\;\;\;\;\frac{{{M_{\sup }}}}{{\Delta t}} < {{\dot M}_{{\rm{ign}}}}}\\
{{{\left( {\frac{{{M_{\sup }}/\Delta t}}{{{M_ \odot }{{\rm s}^{ - 1}}}}} \right)}^{9/4}}\;\;{{\dot M}_{{\rm{ign}}}} < \frac{{{M_{\sup }}}}{{\Delta t}} < {{\dot M}_{{\rm{trap}}}}}\\
{\;{{\left( {\frac{{{{\dot M}_{{\rm{trap}}}}}}{{{M_ \odot }{{\rm s}^{ - 1}}}}} \right)}^{9/4}}\;\;\;\;\;\;\;\;\;\;\;\frac{{{M_{\sup }}}}{{\Delta t}} > {{\dot M}_{{\rm{trap}}}}}
\end{array}} \right\} \ {\rm erg~s}^{-1} \ ,
\end{equation}
where $\chi_{\rm ms}=r_{\rm ms}(a_*)/r_{\rm{g}}$,
$r_{\rm{g}}\equiv 2GM_{\rm BH}/c^2$, $r_{\rm{ms}}$ is the radius of
the marginally stable orbit.

Following the spirit of \citet{Rowlinson2014}, we assume an efficiency $\eta$
as the ratio of the radiated luminosity to the jet power,
where the latter may be regarded as around $L_{\nu \bar\nu}$,
\begin{equation}\label{nuiso}
L_{\rm{X}}=\eta~L_{\nu \bar\nu} \ .
\end{equation}
Here, we adopt $\eta=10\%$ for analyses.
The analytic solution about $r_{\rm{ms}}$
\citep{Bardeen1972} gives $r_{\rm{ms}}=0.97 r_{\rm{g}}$ for
$a_{*}=0.95$, i.e., $\chi_{\rm{ms}}=r_{\rm{ms}}(a_{*})/r_{\rm{g}}=0.97$.
The two critical accretion rates, $\dot{M}_{\rm ign}$ and
$\dot{M}_{\rm trap}$, were given by Equation~(5) of \citet{Zalamea2011}.
We adopt $\alpha=0.1$ as \citet{Zalamea2011} and therefore
$\dot{M}_{\rm ign} \approx 0.021M_{\odot}~\rm{s^{-1}}$
and $\dot{M}_{\rm{trap}} \approx 1.8M_{\odot}~\rm{s^{-1}}$.

\begin{figure}
\figurenum{4}
\centering
\plotone{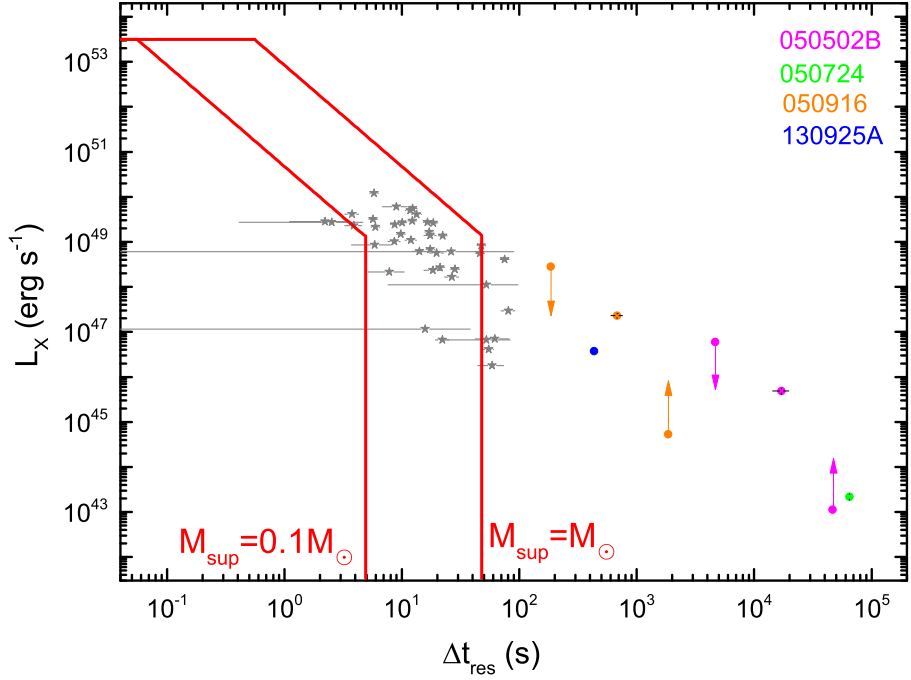}
\caption{
Comparison of the theoretical results (red solid lines) with the observations
(symbols) in the $L_{\rm{X}}-\Delta t_{\rm res}$ diagram for the neutrino
annihilation mechanism.
The theoretical annihilation luminosity is calculated by Equation~(\ref{nunu}).
The left and right red lines correspond to the supplied mass
$M_{\rm sup} = 0.1 M_{\odot}$ and $M_{\odot}$, respectively.
The gray stars are taken from \citet{Luo2013} corresponding to flares
with $\Delta t_{\rm res} < 100$s.
The blue and green circles represent
the two late-time flares with redshift measurement,
whereas the magenta and orange circles represent the other two flares
without redshift measurement, where $z=2$ is adopted.
The lower limit ($z=0.1$) and the upper limit ($z=10$) are also shown by
the arrows in the same colors.
}
\label{F:neutrino luminosity}
\end{figure}

For the neutrino annihilation mechanism, a comparison between the theoretical
results and the observations is shown in Figure~\ref{F:neutrino luminosity}.
For the observations, following the argument in Section~2.2,
we adopt $L_{\rm X} = 0.1 L_{\rm X, iso}$ with regard to
the anisotropic effects.
The red solid lines represent our analytic results for the supplied mass
$M_{\rm sup} = 0.1 M_{\odot}$ and $M_{\odot}$.
Since $M_{\rm sup}$ is the mass supply after the prompt emission,
one $M_{\odot}$ may be regarded as an upper limit for $M_{\rm sup}$.
For a fixed $M_{\rm sup}$, if outflows are not considered,
once the duration $\Delta t_{\rm res}$ is given, the mean mass accretion
rate can be estimated as $\dot M = M_{\rm sup}/\Delta t_{\rm res}$.
Then, by Equation~(\ref{nunu}), we can derive
the theoretical annihilation luminosity.
The gray stars in Figure~\ref{F:neutrino luminosity} represent
the observational results for X-ray flares with $\Delta t_{\rm res} < 100$~\rm{s},
which are taken from \citet{Luo2013}.
The blue and green circles represent
the two late-time flares with redshift measurement,
whereas the magenta and orange circles represent the other two flares
without redshift measurement, where $z = 2$ is adopted.
The lower limit ($z = 0.1$) and the upper limit ($z = 10$)
are also shown by the arrows in the same colors.
As shown in Figure~\ref{F:neutrino luminosity}, the annihilation mechanism
can account for
only a small fraction of X-ray flares with $\Delta t_{\rm res} < 100$~\rm{s}.
As proposed in \citet{Luo2013}, if the magnetic coupling effects between
the inner disk and the central BH are included, the annihilation
mechanism may work for $\Delta t_{\rm res} < 100$~\rm{s}, but still encounter
difficulty in interpreting the flares with $\Delta t_{\rm res} \ga 1000$~\rm{s}.
It is seen from Figure~\ref{F:neutrino luminosity} that,
even for $M_{\rm sup} = M_{\odot}$
the theoretical red line is far from the color circles, which reveals that
the annihilation luminosity is too low to power such late-time X-ray flares.
In addition, other analytic formulae were proposed for the annihilation
luminosity, such as Equation~(11) of \citet{Fryer1999} and Equation~(1)
of \citet{Liu2016}. These two analytic formulae can confirm the above
conclusion that the annihilation mechanism cannot work as the central engine
for the extremely late-time flares with $t_{\rm p} > 10^4$~\rm{s}.

\subsection{Blandford-Znajek process}

An alternative mechanism related to BHs for the central engine
for GRBs and corresponding X-ray flares is the well-known BZ process
\citep{Blandford1977}, where the rotational energy of BHs can be
extracted by the strong magnetic fields and therefore power a
relativistic jet.
Following \citet{Popham1999} and \citet{Di2002},
based on a common assumption that the magnetic field $B$ is around 10\%
of its equipartition value,
the analytic BZ jet power can be expressed as
\begin{equation}
\label{BZ}
P_{\rm BZ} \approx 10^{51} a_*^2
\left( \frac{M_{\rm BH}}{3 M_{\sun}} \right)^2
\left( \frac{\dot M_{\rm in}}{M_{\sun}~{\rm s}^{-1}} \right)
\ {\rm erg~s}^{-1} \ ,
\end{equation}
where $\dot M_{\rm in}$ is the mass accretion rate near the BH horizon.
Similar to the neutrino annihilation case, we use $\eta$ to describe
the ratio of the radiated luminosity to the power,
\begin{equation}\label{BZiso}
L_{\rm BZ}=\eta~P_{\rm BZ} \ ,
\end{equation}
where $\eta=10\%$ is adopted.
For the case of extremely late-time X-ray flares
with duration $\ga 10^4$s,
the mean supplied mass accretion rate ought to be relatively low
$\dot M_{\rm sup} \la M_{\rm sup}/\tau_{\rm res}
\la 10^{-4} M_{\odot}~\rm{s^{-1}}$. In such case,
the neutrino cooling will be negligible compared with the viscous heating.
In this scenario, outflows ought to be quite strong (see the discussions in
Section 4) such that $\dot M_{\rm in}$ will be significantly
less than $\dot M_{\rm sup}$.
Following previous works on outflows, we take the radial profile of
the net inflow accretion rate as
\begin{equation}
\label{MinMout}
\dot M_{\rm in} = \dot M_{\rm out}
\left( \frac{r_{\rm in}}{r_{\rm out}} \right)^p \ ,
\end{equation}
where $p \approx 1$ according to the analyses and simulations of
super-Eddington accretion flows \citep{Ohsuga2005,Gu2012},
and $\dot M_{\rm out}$ may be roughly evaluated as $\approx \dot M_{\rm sup}$.
We assume
$r_{\rm out}/r_{\rm in}\ga 100$ according to the hyper-accretion case.
Then, Equation~(\ref{MinMout}) gives $\dot M_{\rm in}/\dot M_{\rm out} \la 1\%$.
In other words, more than 99\% of the supplied mass will not enter the BH,
but escape from the disk by outflows. It is obvious from Equation~(\ref{BZ})
that such outflows will have essential influence on the power of the
BZ process.
By combining Equations~(\ref{BZ})-(\ref{MinMout})
we can derive the following relationship:
\begin{equation}
\label{Lx}
L_{\rm X} \approx 10^{48} a_*^2 \left(
\frac{M_{\rm sup}/\Delta t_{\rm res}}{M_{\sun}~{\rm s}^{-1}} \right)
\ {\rm erg~s}^{-1} \ .
\end{equation}

In order to directly compare the theory and the observation,
the anisotropic effects of jet radiation should be considered.
The relation between the BZ luminosity and the isotropic
luminosity takes the form,
\begin{equation}
\label{theta}
L_{\rm BZ} = L_{\rm X,iso} \ (1-\cos\theta_{\rm jet}) \ ,
\end{equation}
where $1-\cos\theta_{\rm jet} \approx 0.1$ is adopted as discussed in
Section~2.2.

A comparison of the theoretical results with the observations is shown in
Figure~\ref{F:BZ luminosity}.
The two pairs of solid lines correspond to our theoretical
results for $a_* = 0.8$ (green) and 0.95 (red), which are
calculated by Equation~(\ref{Lx}).
For the two lines in the same color, the upper line corresponds to
$M_{\rm{sup}}=M_{\odot}$, and the lower line corresponds to
$M_{\rm{sup}}=0.1M_{\odot}$.
Similar to Figure~4, the blue and green circles represent the observed
X-ray flares with redshift measurements.
The magenta and orange circles represent the
other two flares without redshift measurement, where $z=2$ is adopted.
The lower limit ($z=0.1$) and the upper limit ($z=10$) are also shown
by the arrows in the same colors.
It is seen from Figure~\ref{F:BZ luminosity} that, even for the extreme case
with $M_{\rm sup}=M_{\odot}$ and $a_* = 0.95$, the BZ mechanism can hardly
account for these four flares.
The physical reason is that, due to the strong outflows,
the mass rate near the BH horizon is quite low
(generally $\la 10^{-6} M_{\odot}~{\rm s^{-1}}$), and therefore
the magnetic fields accumulated by the accretion will probably
be too weak to power such flares.
Thus, for the extreme late-time X-ray flares,
the BZ mechanism may not work as the central engine either.

As shown by Equation~(\ref{BZ}), we adopt $L_{\rm BZ} \propto a_*^2$ for
the analyses, which agrees with the simulation results on geometrically
thin disks \citep{Tchekh2010a}. On the other hand, \citet{Tchekh2010a}
showed that, for geometrically thick disks, however, the relation
will be $L_{\rm BZ} \propto a_*^4$ or even $\propto a_*^6$ for fast
rotating cases ($a_* \to 1$). In other words, the jet luminosity
$L_{\rm BZ}$ will decrease sharply with decreasing $a_*$
\citep[e.g., Figure~6 of][]{Tchekh2010a}.
For the flow with relatively low accretion rates
$\la 10^{-4} M_{\odot}~{\rm s^{-1}}$,
the neutrino cooling will be inefficient and therefore the disk is likely
to be geometrically thick \citep{Gu2015}.
As a consequence, our analytic $L_{\rm BZ}$ for $a_* < 1$ may be overestimated
and therefore $L_{\rm BZ}$ in the real cases may be even lower such that
it may not be responsible for the late-time flares.

\subsection{Neutron star with strong bipolar magnetic fields}

The above two subsections have shown that the central engine for
the extremely late-time X-ray flares is unlikely
to be associated with BHs.
In this subsection, we study
the possibility of an NS system as the engine.
Such a model invokes a rapidly spinning,
strongly magnetized NS or a magnetar
\citep{Usov1992,Thompson1994,Dai1998,Wheeler2000,Zhang2001,Metzger2008,
Metzger2011,Bucciantini2012}.
In the NS scenario, the energy reservoir is the total rotational energy of
the magnetized NS, (see Equation~(1) in \citet{Lu2014L}):
\begin{equation}
\label{rot energy from NS}
E_{\rm {rot}}\approx 2\times 10^{52} \frac{M_{\rm NS}}{1.4M_{\sun}}
\left( \frac{R}{10^{6}{\rm cm}} \right)^{2}
\left( \frac{P_0}{10^{-3}{\rm s}} \right)^{-2}
\ {\rm erg} \ ,
\end{equation}
where $M_{\rm NS}$ is the NS mass and $P_0$ is the initial spin period.
We therefore take $E_{\rm rot} = 2\times10^{52}${\rm erg} as a typical
rotational energy.
Another energy source is the magnetic energy of a magnetar.
The total magnetic energy in a magnetar can be roughly calculated by
\begin{equation}
E_{\rm {mag}}\approx \frac{B^{2}}{8\pi} \times \frac{4}{3}\pi R^{3}
= \frac{1}{6}B^{2}R^{3} \ ,
\end{equation}
where $B$ is the poloidal magnetic field strength on the horizon.
With $R\approx 10^{6}\rm{cm}$, the total magnetic energy is
$E_{\rm mag} = 1.7\times 10^{47} {\rm erg}$ for $B=10^{15}{\rm G}$.

A comparison of the rotational energy and the magnetic energy
with the observations of the X-ray flares is shown in Figure~\ref{F:NS energy}.
Again, we choose $1-\cos\theta_{\rm jet} = 0.1$ due to
the anisotropic effects, and the radiative efficiency $\eta = 10\%$.
For a GRB with multiple flares, we plot the total energy of all the flares
instead of the single late-time flare.
It is seen that, the energy of flares is significantly larger than
the magnetic energy even for $B=10^{15}$G,
which implies that the magnetic energy may not power the X-ray flares.
Or, the magnetic energy can only work with extremely strong magnetic
fields $B \gg 10^{15}$G.
Such an issue has been investigated by \citet{Dai2006}.
On the other hand, the rotational energy (red dashed line)
is obviously higher than all the observational data, which indicates that
the rotational energy may be responsible for
the late-time X-ray flares in our sample.

\section{Conclusions and discussion}

\begin{figure}
\figurenum{5}
\centering
\plotone{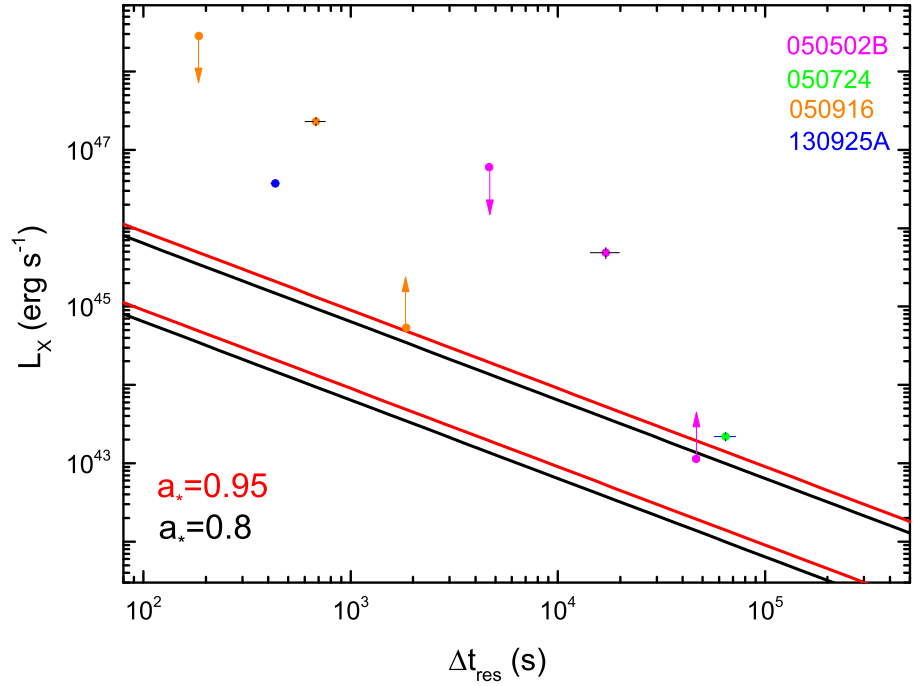}
\caption{
Comparison of the theoretical results of the BZ mechanism with
the observations, where two typical values for the spin parameters
$a_{*} = 0.8$ (black) and 0.95 (red) are fixed.
For the two lines in the same color, the upper line corresponds to
$M_{\rm sup} = M_{\odot}$, and the lower line corresponds to
$M_{\rm sup} = 0.1~M_{\odot}$.
The meaning of the circles and arrows is the same as in
Figure~\ref{F:neutrino luminosity}.
}
\label{F:BZ luminosity}
\end{figure}

\begin{figure}
\figurenum{6}
\centering
\plotone{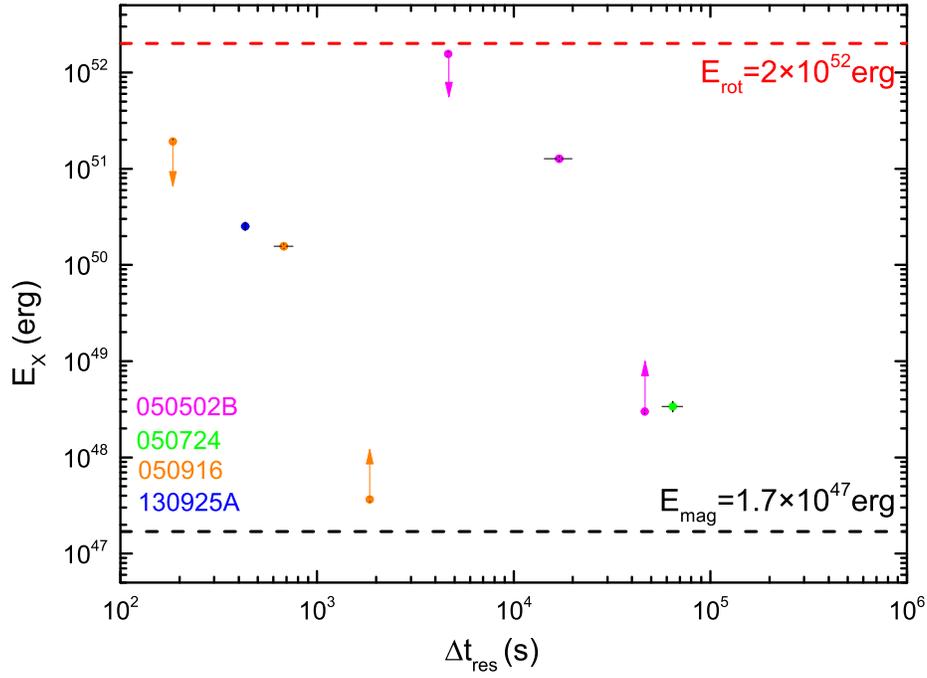}
\caption{
Energy of the X-ray flares in our sample.
For a GRB with multiple flares, the total energy of all the flares is plotted.
The red dashed line represents the energy related to the
rotational energy of a fast rotating NS,
and the black dashed line corresponds to the energy related to
the magnetic energy of a magnetar with $B=10^{15}{\rm G}$.
The meaning of the four colors
is the same as in Figure~\ref{F:neutrino luminosity}.
}
\label{F:NS energy}
\end{figure}

The present work focuses on the central engine of extremely late-time X-ray
flares ($t_{\rm p} > 10^4 {\rm s}$) with the internal origin assumption.
We have investigated the possibility of the two well-known mechanisms
related to BHs for the central engine, i.e., the neutrino-antineutrino
annihilation and the BZ process.
Our results show that the annihilation luminosity is far below
the observational data, which indicates that the annihilation mechanism
cannot account for the extremely late-time X-ray flares.
On the other hand, for the BZ process, if the role of outflows is
taken into consideration, the inflow mass rate near the horizon
will be quite low such that the magnetic field will probably be too weak
to power the observed X-ray flares.
We therefore argue that, for such late-time X-ray flares,
the central engine is unlikely to be associated with BHs.
On the contrary, a fast rotating NS with strong bipolar
magnetic fields may be responsible for such flares.
We would stress that this work only considered bright flares.
Some dim flares at late-time may be missed since the underlying continuum
is too bright for their detection. These dim flares may occupy the
lower part of Figures~\ref{F:BZ luminosity} and \ref{F:NS energy}, which are possibly consistent
with the BZ mechanism and the magnetic origin in the magnetar context
\citep{Margutti2011}.

In this work, the existence of outflows is a key point to draw the conclusion
that the BZ mechanism is unlikely to power the extremely late-time
X-ray flares. In recent years, outflows have been found
to be significant in accretion systems of different scales by
theories \citep[e.g.,][]{Jiao2011,Gu2015}, simulations
\citep[e.g.,][]{Ohsuga2005,Ohsuga2011,Yuan2012a,Yuan2012b,Jiang2014,
Sadow2015,Sadow2016},
and observations \citep[e.g.,][]{Wang2013}.
Based on the balance of heating and cooling, \citet{Gu2015} shows that
the outflow is inevitable for the accretion flows that the radiative cooling
is far below the viscous heating, no matter the flow is optically thin
or thick. In the current work for accretion rates
$\dot M \la 10^{-4} M_{\sun} {\rm s}^{-1}$, neither the photon radiative
cooling nor the neutrino one is efficient to balance the viscous heating.
Thus, the outflows ought to be significant. Actually, \citet{Liu2008}
studied this issue and proposed that there exists a lower critical $\dot M$
varying with radius, below which outflows have to occur.
From the observational view, taking our Galactic center as an example,
\citet{Wang2013} reveals that more than 99\% of the accreted mass escape
from the accretion flow by outflows.
Therefore, it is reasonable to
assume less than 1\% of the supplied mass can enter the BH
in the present work.

The present work focuses on the late-time X-ray flares with
$t_{\rm p} > 10^4 {\rm s}$. On the other hand,
a previous work \citep{Luo2013} focused
on the X-ray flares with rest duration $\Delta t\la 100{\rm s}$,
and found that the neutrino annihilation mechanism cannot account for
the flares except for including the magnetic coupling between the inner
disk and the BH. However, such a coupling and corresponding distribution
of magnetic fields have not been found in simulations yet. Thus,
we would argue that, in general, the annihilation mechanism may not work
as the central engine for X-ray flares. For the BZ mechanism,
the output power is larger than that of the annihilation mechanism,
particularly for relatively low accretion rates. From the energy argument,
the BZ mechanism may be responsible for X-ray flares with duration
$\Delta t_{\rm res} \la 10^4~{\rm s}$ in the case that outflows
are not significant.

\acknowledgments

We acknowledge the use of the public data from the Swift data archive.
We thank Xue-Feng Wu, Hao Tong, Bing Zhang, and Jirong Mao
for beneficial discussions,
and thank the referee for constructive suggestions that improved the paper.
This work was supported by the National Basic Research Program of China
(973 Program) under grants 2014CB845800,
the National Natural Science Foundation of China under grants 11573023,
11533003, 11503011, 11473022, 11403005, 11333004, 11233006, 11222328,
and U1331101,
the CAS Open Research Program of Key Laboratory for the Structure and
Evolution of Celestial Objects under grant OP201503,
and the Fundamental Research Funds for the Central Universities
under grants 20720140532 and 20720160024.

\clearpage

\clearpage

\clearpage

\clearpage

\clearpage

\clearpage

\clearpage

\floattable
\begin{deluxetable}{lcccccccc}
\rotate
\centering
\tablewidth{0pt}
\tabletypesize{\footnotesize}
\tablecaption{Fitting results of the late-time flares.}
\tablecolumns{9}
\tablenum{1}
\tablehead{
\colhead{$\rm{GRB}$}
& \colhead{$z$}
& \colhead{$\alpha_{1}$}
& \colhead{$\alpha_{2}$}
& \colhead{$t_1-t_2$~(10$^4$ s)}
& \colhead{$t_{\rm p}$~(10$^4$ s)}
& \colhead{$F_{\rm p}~(\rm{erg~cm^{-2}s^{-1}})$}
& \colhead{$\beta$}
&\colhead{$\chi^{2}/dof$}
}
\startdata
050502B	&	--	&	-1.40 	$\pm$	0.41 	&	5.56 	$\pm$	1.18 	&	3.22 	-	12.82 	&	7.63 	$\pm$	 0.38 	 &(	 1.08 	$\pm$	0.15 	)$\times$	$10^{-12}$	&	$	0.95 	^{+	0.25 	}_{	-0.11 	}$&
2.33\\
050724	&	0.257	&	-2.04 	$\pm$	0.44 	&	3.66 	$\pm$	0.52 	&	1.72 	-	22.09 	&	5.56 	$\pm$	 0.31 	 &(	1.90 	$\pm$	0.22 	)$\times$	$10^{-12}$	&	$	0.59 	^{+	0.35 	}_{	-0.22	}$&
1.32	\\
050916	&	--	&	-33.68 	$\pm$	3.10 	&	11.20 	$\pm$	1.80 	&	1.68 	-	3.06 	&	1.93 	$\pm$	 0.02 	 &(	 3.76 	$\pm$	1.15 	)$\times$	$10^{-11}$	&	$	0.53 	^{+	0.22 	}_{	-0.17 	}$&1.27	\\
130925A	&	0.347	&	-24.49	$\pm$	2.01 	&	57.02 	$\pm$	1.11 	&	 1.01 	-	1.19 	&	1.12 	$\pm$	 0.01 	 &(	6.18 	$\pm$	0.19 	)$\times$	 $10^{-10}$	 &	$	1.34 	^{+	0.34 	}_{	-0.31	}$&
1.62	\\
\hline
070318	&	0.84	&	-1.52 	$\pm$	1.28 	&	2.84 	$\pm$	1.43 	&	 14.03 	-	37.21 	&	19.46 	$\pm$	 3.02 	 &(	3.35 	$\pm$	1.08 	)$\times$	 $10^{-13}$	 &	$	0.79 	^{+	0.37 	}_{	-0.34	}$& 1.53	\\
070429A	&	--	&	-11.54 	$\pm$	11.72 	&	2.31 	$\pm$	0.61 	&	21.23 	 -	 90.79 	&	23.44 	$\pm$	 1.43 	&(	 2.13 	$\pm$	0.05 	)$\times$	$10^{-13}$	&	 $	1.23 	^{+	0.44 	}_{	-0.40	}$& 1.15	 \\
150626B	&	--	&	-1.08 	$\pm$	0.12 	&	2.23 	$\pm$	0.11 	&	0.33 	-	114.37	 &	 2.11 	$\pm$	 0.08 	&(	 1.08 	$\pm$	0.05 	)$\times$	$10^{-13}$	&	$	1.03 	^{+	0.21 	}_{	-0.20 	}$& 0.86	\\
\enddata
\end{deluxetable}

\clearpage

\floattable
\begin{deluxetable}{lccccc}
\rotate
\tablecaption{Physical parameters based on the fitting results}
\tablehead{
\colhead{$\rm{GRB}$}
&\colhead{$\Delta t~({\rm s})$}
&\colhead{$\Delta F/F$}
&\colhead{$E_{\rm{X}}~(\rm{erg})$}
&\colhead{$L_{\rm{X}}~(\rm{erg~s^{-1}})$}
&\colhead{$E_{\rm{X,all}}~(\rm{erg})$}
}
\centering
\tablewidth{0pt}
\tabletypesize{\footnotesize}
\tablecolumns{6}
\tablenum{2}
\startdata
\hline
050502B	&(	5.11	$\pm$	0.82	)$\times$	$10^{4}$ 	&	6.64 	$\pm$	0.92 	&(	8.30	 $\pm$	 0.27	 )$\times$	$10^{49}$ 	&(	4.87	$\pm$	0.79	)$\times$	$10^{45}$ 	&(	1.27 	 $\pm$ 0.41 	 )$\times$	 $10^{51}$ 	\\
050724	&(	8.11 	$\pm$	0.97	)$\times$	$10^{4}$ 	&	29.38 	$\pm$	3.40 	&(	1.42	 $\pm$	 0.15	 )$\times$	$10^{48}$ 	&(	2.19	$\pm$	0.35	)$\times$	$10^{43}$ 	&(	3.39 	 $\pm$	0.42 	 )$\times$	 $10^{48}$ 	\\
050916	&(	2.04	$\pm$	0.23	)$\times$	$10^{3}$ 	&	87.70 	$\pm$	26.96 	&(	1.56	 $\pm$	 0.07	 )$\times$	$10^{50}$ 	&(	2.30	$\pm$	0.28	)$\times$	$10^{47}$ 	&	--			 	 \\
130925A	&(	0.58	$\pm$	0.03	)$\times$	$10^{3}$ 	&	15.71 	$\pm$	0.48 	 &(	1.62	$\pm$	0.06	 )$\times$	$10^{49}$ 	&(	3.74	$\pm$	0.22	)$\times$	$10^{46}$ 	 &(	2.52 	$\pm$	0.06 	)$\times$	 $10^{50}$	\\
\hline
070318	&(	8.71	$\pm$	5.41	)$\times$	$10^{4}$ 	&	3.96	$\pm$	0.63	&--	&--	 &--	 \\
070429A	&(	8.65	$\pm$	2.12	)$\times$	$10^{4}$ 	&	3.28    $\pm$	0.78	&--	&--	 &--	 \\
150626B	&(	1.18	$\pm$	0.18	)$\times$	$10^{4}$ 	&	50.63	$\pm$	2.34	&--	&--	&--	\\
\enddata
\end{deluxetable}

\end{document}